\documentclass[a4,11pt]{article}
\usepackage[margin=2.5cm]{geometry}
\usepackage[affil-it]{authblk}
\usepackage{footmisc}
\usepackage[hidelinks]{hyperref}
\usepackage{graphicx}
\usepackage{amsmath}

\title{\textbf{Unveiling music genre structure through common-interest communities}}

\author[1]{Zhiheng Jiang}

\author[2,*]{Hoai Nguyen Huynh}

\affil[1]{Hwa Chong Institution, Singapore}

\affil[2]{Institute of High Performance Computing, Agency for Science, Technology and Research, Singapore\vspace{0.3cm}}

\affil[*]{\textnormal{\small\textbf{Corresponding author:} \texttt{huynhhn@ihpc.a-star.edu.sg}}}

\date{}

\begin{document}

\maketitle

\abstract{Using a dataset of more than $90,000$ metal music reviews written by over $9,000$ users in a period of $15$ years, we analyse the genre structure of metal music with the aid of review text information. We model the relationships between genres using a user-oriented network, based on the written reviews. We then perform community detection and employ a network ``averaging'' method to obtain stable genre clusters, in order to analyse the structures of clusters both locally within each cluster and globally over the entire network. In addition to identifying the clusters, we use Dependency Parsing and modified Term Frequency - Inverse Document Frequency to extract significant and unique features of each cluster. These structures and review text information can allow us to understand how music audience (fans) perceive similar and different genres, and also assist in classifying different genres which share common-interest user communities, offering a more objective way in grouping music genres. Furthermore, the classification can also help recommendation engines provide more targeted suggestions of music, and potentially help musicians to select genre labels for their music, and design music to better cater to preferences of their audiences based on previous reviews.}

\vspace{0.2cm}\textbf{Keywords:} Complex networks, Social networks, Community detection, Music genre, Metal music, Text mining

\section{Introduction}\label{sec:introduction1}

The availability of music-related data in digital format has allowed deeper understanding of the music content per se as well as the relationship between music and its target audience. Among the available data, review text is a useful source as it can be mined to obtain trends and patterns encoded in the music. Such review data analysis is useful in investigating the genre structure of music, which is often difficult to describe, due to the wide range of criteria involved, and the tendency of genres to evolve over time \cite{bazerman_describe_2009}. While genres are usually subjective and determined (or labelled) by those who listen to the music (such as reviewers) at individual level, a global approach (i.e. a collective user-oriented approach) to combine the individual musical preferences can produce an objective way to classify these genres. The unique review patterns of users can reveal the common perceptions of how different genres are related, not by their name but the music content therein. In that context, the object of this research is twofold, aiming to investigate both the music genre structures through common-interest communities and the effectiveness of using review text analysis to understand the genre structures.

Music genres are categories which describe different types of music with certain similar characteristics, based on the nature of music, as well as the people who listen to these types of music. In other words, genres are used to group together both music as well as the people who listen to particular types of music \cite{roy_what_2010}, or common-interest communities. Genre boundaries are important as they set expectations for people who listen to music, and also serve as a reference when musicians create music for their intended audiences \cite{walser_running_2014}. Genre classification, therefore, is a popular area for research, largely due to the rapid growth of the music industry. As the number of albums and genres increases, genre classification has become increasingly important for organisation of large music databases \cite{tsatsishvili_automatic_2011}. For music fans, genre classifications can also be a form of social and political identity \cite{lena_2012}, and it can help listeners to find similar genre communities and listen to the music they like \cite{dimaggio_classification_1987}. Music is often a form of self expression, from which we can observe patterns of social class and cultural habits based on musical tastes and preferences \cite{coulangeon_social_2005,coulangeon_testing_2005}. Past research based on multiple countries has shown that music classifications are often related to social categories, such as gender, and can show certain hierarchical structures \cite{schmutz_social_2009}. Analysing these music genres can also reveal certain stereotypes about genres, which often can reflect accurate information about personality \cite{rentfrow_content_2007}. In that perspective, music review written by the music receivers can be used to extract musical characteristics, or perceptions of certain types of music, for a collection of similar genres.

Furthermore, analysing genre structures is useful in helping us understand how genres evolve over time. Over time, existing genres will evolve and creates genre hierarchies, where a main genre expands into further subgenres. One of the reasons for the increasing number of subgenres is that over time, there is greater diversity due to the introduction of new music styles, and fusion between different music styles. This creates the need for more ways to describe different subgenres. In addition, it is also a marketing technique for music organisations to appeal to the consumer culture today \cite{mcleod_genres_2001}.

In the literature, existing methods mostly focus on clustering music fans, bands or artists as nodes in a network \cite{makkonen_north_2017, oliveira_musical_2020, sokolov_is_2018}, for example in the form of collaboration networks. While such networks are useful in describing the relationships between different music fans or musicians, they are less useful when it comes to genre analysis, as compared to genre networks. Genre networks have been used in past research to explain the relationships between different genres in varying cultural contexts, such as book genres and music genres. These genre networks are often constructed based on genre relationships which are dependent on the books \cite{muntea_network-based_2020,sokolov_is_2018}, albums \cite{silver_genre_2016}, lyrics \cite{atherton_i_2016} or shared CDs \cite{park_topology_2015} that different genres have in common, hence forming the links between different genres. Other genre networks have been constructed using the similarity of the music itself, such as using music scores \cite{correa_finding_2011} or audio files \cite{tsunoo_audio_2009}. However, these research do not take into account that genres revolve around the people who listen to these types of music, whose preferences are ultimately the reason behind why genres are differentiated in a particular way.

There has been limited research on user-oriented genre networks, such as \cite{lizardo_mutual_2018}, which was conducted based on survey data and only subdivided the music genres into two communities. In addition, this research had low sample size of only 20 genres, which made it difficult to decipher much meaningful information from the community detection. Moreover, after identifying different genre communities, it remains difficult to make sense of the reasons why genres are clustered in the same community, and what common characteristics they share, especially for larger genre networks.
Along this line, Music Information Retrieval (MIR) is an interdisciplinary field which involves extracting useful information in music based on information such as musical characteristics (e.g. rhythm, pitch and lyrics) \cite{downie_music_2003}. However, the problem with MIR is that it does not take into account that genre concepts can vary from person to person. Furthermore, genre boundaries can shift over time due to changing tastes and preferences, which makes training genre classification models problematic. Hence, an approach that takes into account the subjectivity and flexibility of genres such as review-based classification method will be more appropriate in classifying the music genres.

Coming from the receiving end, reviews are a form of music criticism or positive feedback which has been shown to be the main driver of discourse and evolution of music genres, and greatly influences how music is created and received \cite{alessandri_consumers_2020}. Typical music reviews comprise of a few paragraphs, explaining why the author likes or dislikes a particular album. The reviews offer valuable information about the album and the genre itself, as these reviews often contain key features of a particular genre which are enjoyed by fans, such as vocal, musical or lyrical characteristics. This kind of information can be extracted to provide insights into the clusters of genres that we aim to study.

In this paper, a user-oriented approach is proposed to identify similar music genres based on common-interest communities, and to analyse the structures and characteristics of the genre clusters formed, with reference to the review text. This is because genre classification is often subjective in nature, and classifying them based on common users is sensible as the common users form common-interest communities which listen to related genres. Metal music was chosen in this study because it is known to be starkly different from mainstream music as it carries an attitude of anti-commercial rewards that is widely shared by both the artists and the fans \cite{weinstein_heavy_1991,moynihan_lords_1998}. In some instances, the attitude is taken to extreme and becomes a philosophy \cite{moynihan_lords_1998}, often due to the conservativeness of metal music fans, which would likely engrave distinctive patterns in the clusters of music genres. There have been cases where a formerly well-regarded band decided to turn commercial (e.g. producing 'easy-to-listen' songs, signing with commercial label records etc.) and has been detested by its own fans \cite{christe_sound_2003}. Having said that, the community of fans in metal music appears to possess unique characters that are different from those of other types of music \cite{arnett_three_1993,weinstein_heavy_2009}, ranging from dislike of authority, desire to be unique \cite{swami_etal_2013}, to strong loyalty\footnote{\url{https://loudwire.com/spotify-study-metal-fans-most-loyal/}}. Therefore, the subgenres derived from metal music have their own unique characteristics and are capable of acting as almost independent cultural objects \cite{lena_2012}.

With all of the above, the main hypothesis being put to the test in this paper is that similar genres (i.e. subgenres of metal music) will likely be listened to by a similar community of users, manifesting genre structures, and that user reviews contained in each cluster of genres are likely to have unique text features compared to other clusters. Specifically, this paper aims to uncover the latent subdivision of metal music genres through objective clustering based on common-interest communities, which is meant to be more organic and less subjective. After identifying the clusters of genres, the paper aims to analyse the structure of the clusters obtained, and thereafter uses Natural Language Processing techniques to extract key album features from reviews in order to analyse the characteristics of these genre clusters and draw insights from the clusters obtained.

\section{Data and methodology}

\subsection{Data}

The data in this study was obtained from Encyclopedia Metallum - Metal Archives\footnote{\url{https://www.metal-archives.com/}} (hereafter abbreviated as MA), a comprehensive database of metal albums and reviews, and an essential source of information for metal fans. This database was used to obtain the user and album information, as well as reviews. Details of the data can be found in Table \ref{tab:description}.

\begin{table}[t]
\caption{Description of data from the Metal Archives.} 
\label{tab:description}
\centering 
\begin{tabular}{c c} 
\hline\hline 
Variable & Value \\ [0.5ex] 
\hline 
Time period of reviews & July 2002 -- June 2017 \\
Time period of albums & 1968 -- 2017 \\
Number of reviews & 93,434 \\ 
Number of users & \hphantom{0}9,317 \\
Number of albums & 40,241 \\
Number of bands & 20,574 \\
Average length of review & 513 words \\ [1ex] 
\hline 
\end{tabular}
\end{table}

All reviews submitted and accepted to MA over a period of 15 years (from July 2002 to June 2017) were collected for this study. The reviews were written for 40,241 albums released between 1968 and 2017. Each review contained a score, out of 100, and also accompanying text describing each album. The average length of each review was around 500 words. The genre information was mostly obtained from Metal Storm\footnote{\url{http://www.metalstorm.net/home/}} (MS), a separate database for metal music, to complement the genre data on MA. MS was used as MA only had genre information for bands, but not individual albums, unlike MS. Examples of the genres are shown in Table \ref{tab:topgenres}, listing 20 genres that received the most number of reviews.

\begin{table}[t]
\caption{Top genres sorted by number of reviews (the numbers in the brackets are for reviews with score 75 and above, see Sec. \ref{sec:bipartite_network})}
\label{tab:topgenres}
\centering 
\begin{tabular}{c c c c} 
\hline\hline 
Rank & Genre & Number of reviews & Number of users \\ [0.5ex] 
\hline 
1 & Black metal & 14887 (9948) & 3056 (2722) \\
2 & Death metal & 11007 (7783) & 2543 (2271) \\
3 & Heavy metal & 9473 (6536) & 2038 (1722) \\
4 & Thrash metal & 7426 (5124) & 1812 (1580) \\
5 & Power metal & 5033 (3547) & 1277 (1096) \\
6 & Progressive metal & 3482 (2544) & 1280 (1105) \\
7 & Doom metal & 3166 (2379) & 1089 (976) \\
8 & Melodic death metal & 2672 (1944) & 1110 (956) \\
9 & Brutal death metal & 1908 (1389) & 784 (659) \\
10 & Technical death metal & 1576 (1187) & 753 (649) \\
11 & Folk metal & 1536 (1122) & 681 (586) \\
12 & Atmospheric black metal & 1493 (1095) & 615 (539) \\
13 & Grindcore & 1490 (1077) & 683 (567) \\
14 & Gothic metal & 1362 (971) & 637 (527) \\
15 & Symphonic metal & 1148 (752) & 547 (408) \\
16 & Death doom metal & 1124 (799) & 545 (459) \\
17 & Sludge metal & 1115 (825) & 532 (461) \\
18 & Symphonic power metal & 1056 (753) & 466 (378) \\
19 & Stoner metal & 1050 (806) & 481 (402) \\
20 & US power metal & 1006 (728) & 395 (322) \\
\hline 
\end{tabular}
\end{table}

\subsection{Methodology}

We adopted a two-pronged approach to analyse genre communities. Firstly, we constructed a bipartite graph between genres and users, and removed some of the users who have reviewed a large proportion of genres. Thereafter, we constructed the network of genres, and applied further pruning using $k$-core algorithm, before performing a modified version of community detection using modularity maximisation, for multiple rounds, to obtain our genre communities. We then analysed the structural characteristics of these communities both at the network and cluster level. For each genre cluster, we also extracted the distinctive features from the relevant review texts pertaining to each genre in the genre cluster, using Dependency Parsing and Term Frequency-Inverse Document Frequency, to explain the qualitative aspects of the genre clusters. Hence, we are able to describe the genre clusters based on both structural and review-based characteristics.

\subsubsection{Construction of bipartite graph between genres and users}
\label{sec:bipartite_network}

\begin{figure}[t]
\centering
\includegraphics[width=0.7\textwidth]{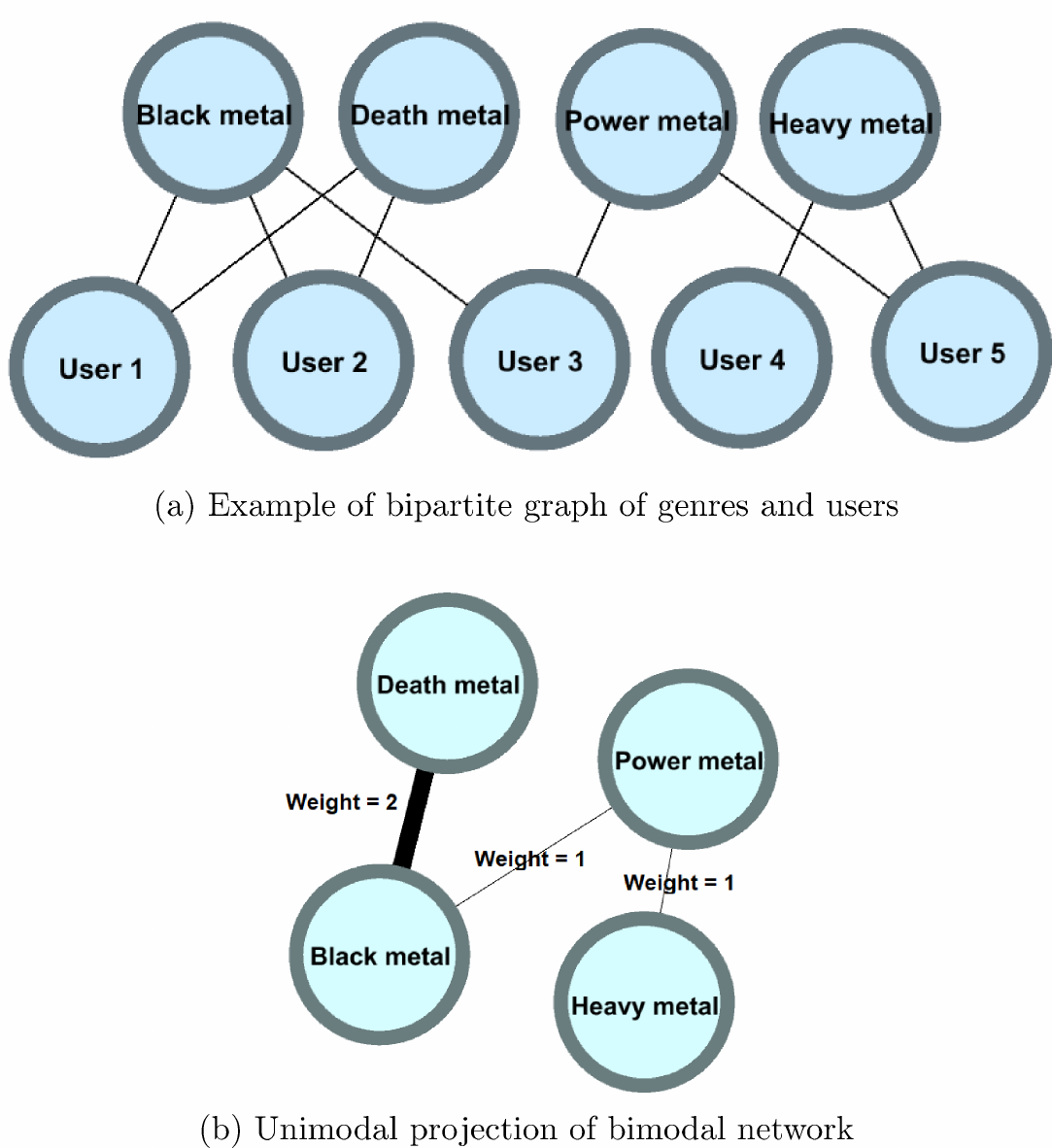}
\caption{\label{fig:bipartite}Network of genres and users. The edge weight in the unimodal network reflects the number of distinct users connecting the genres in the bimodal network.}
\end{figure}

\begin{figure}[t]
\centering
\includegraphics[width=\textwidth]{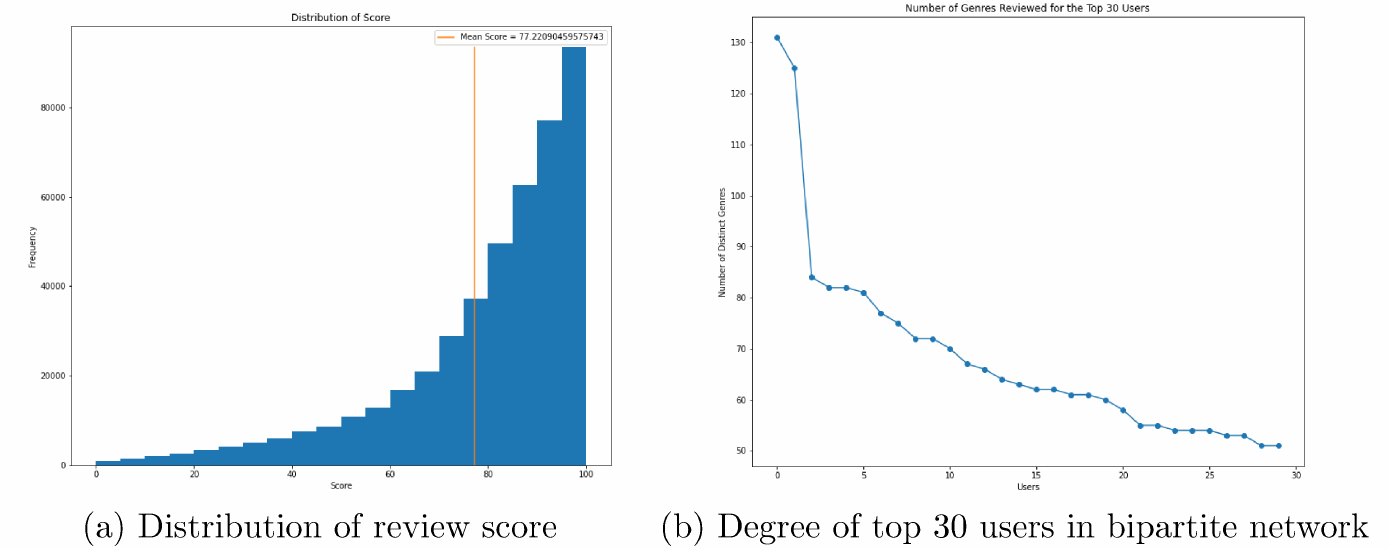}
\caption{\label{fig:data_stat}Statistics of review data.}
\end{figure}

With each album tagged with a corresponding set of genres, we constructed a bipartite graph (i.e. bimodal network) between users and genres (see Fig. \ref{fig:bipartite}). An edge is constructed in the bimodal network from a genre to a user if that user has positively reviewed an album of that particular genre before. A positive review score is defined to be a score of at least 75 out of 100, deduced based on the score distribution (see Fig. \ref{fig:data_stat}a). The mean score was found to be 77, which is close to the chosen score threshold. We also tested with varied score threshold between 70 and 80 and found no significant difference in the clustering outcome.

While the positive reviews indicate that the reviewers enjoy the albums (and likely the genres), and hence, can be used to make connections between the genres, the negative ones do not necessarily play an equal role. From our inspection of the negative reviews, the low scores are given for different reasons, including the reviewers not enjoying the genres, low music quality, lack of musical novelty, high expectation of the reviewers, etc. In many cases, reviewers who enjoy previous albums of a band of a certain genre tend to give very low score for new albums in which the band change their genre, indicating the reviewers' disinterest in that genre, which does not make a genuine link between the reviewers and the genre. For this, we found that the negative reviews added more noise than useful information, and decided to exclude them from the analysis.

\subsubsection{Filtering unusual users}

Certain users are found to influence the structure of the network to a much greater extent than others, by reviewing a huge proportion of the genres. One explanation could be that these users have eclectic music tastes, and enjoy a wide variety of genres. One example of this is people who are cultural omnivores, and have a diversity of genre preferences which often cut across different social classes \cite{coulangeon_testing_2005,cook_cultural_2015,warde_understanding_2007}.  This resulted in a denser network of genres as more non-related genres were connected together by these users. By removing these users who have reviewed a large number of genres, fewer genres are unnecessarily connected to each other in the network, and unrelated genres are less strongly connected compared to before. As shown in Fig. \ref{fig:data_stat}b, there is a sharp drop in the number of genres positively reviewed in the graph after the top two users with highest number of genres reviewed. It is reasonable to assume that these rare omnivores do not well represent the general tastes and preferences of the metal music community. Hence, these two users were removed before the network of genres was constructed, and their removal should not break any reasonable connections that majority of the users believe them to be. In other words, if many other users actually re-establish the removed connections, they will be recovered in the network.

\subsubsection{Construction of unimodal network of genres from bipartite network}

The bimodal network is then reduced to a unimodal network of genres using a similar method as \cite{padron_alternative_2011}. If two different genres are both positively reviewed by at least a single user, an edge is drawn between these two genres. If an album has multiple genres, the genres of that album will also be connected as the user is taken to have reviewed all genres of that album. The weight of the edge between two genres, e.g. genre A and genre B, is the number of distinct users who have (positively) reviewed both genres A and B. In other words, we can use the bipartite graph constructed earlier to determine the number of users shared by both genres. An example of the resultant unimodal network is shown in Fig. \ref{fig:bipartite}b. In other words, two genres are related to each other if common users have positively reviewed both genres. As this network of genres is constructed based on users as edges between different genres, this network is user-oriented. 

\subsubsection{Removing genres using $k$-core algorithm} 
The $k$-core decomposition method was used to remove noisy peripheral genres which lack enough data for analysis. Instead of the common procedure of removing edges based on low edge weight \cite{yan_weight_2018}, the $k$-core algorithm was performed on the network of genres. The $k$-core algorithm starts with removing nodes (genres) which have degree of 1 repeatedly, until there are no more nodes with degree $1$. Then the algorithm proceeds to remove nodes of degree $2$, $3$, and until degree $k$. In other words, the $k$-core of a graph (or network) $G$ is the subgraph obtained after continuously removing nodes in the network such that every node remaining in the network has at least degree $k$ \cite{khaouid_k-core_2015}. This algorithm terminates when the main core of the network is obtained, in other words, when the value of $k$ becomes the maximum. This effectively removed peripheral genres which are more likely to act as noise due to the lack of data. Before the $k$-core decomposition, we had 190 genres, and this number was reduced to 109 genres contained in one main core after the $k$-core decomposition. After evaluating the genres removed, we determined that most of the insignificant genres were removed since these genres had low node degree.  This initial filtering of nodes proved to improve the clusters obtained later on.

\subsubsection{Community detection}
To partition the network into different distinct clusters, the Louvain method for community detection \cite{blondel_fast_2008} was used. This method aims to maximise the modularity score of a type of partitioning, and uses a greedy algorithm to extract communities from large networks. The community detection algorithm uses the fact that nodes belonging to different communities will be relatively sparsely connected to each other, while nodes belonging in the same community are more densely connected to each other. Modularity refers to a score given to a way of partitioning of a particular network, which reflects the quality of the partitions. As optimisation of modularity is very computationally intensive, approximation algorithms are necessary. The method of community detection by \cite{blondel_fast_2008} finds high modularity partitions in a short amount of time, making it one of the most efficient community detection algorithms available. We found that this method of community detection was also advantageous as it yielded more regular cluster sizes, as compared to other clustering algorithms like $k$-means clustering \cite{likas_global_2003}.

\subsubsection{Average network partitioning}

\begin{figure}[t]
\centering
\includegraphics[width=\textwidth]{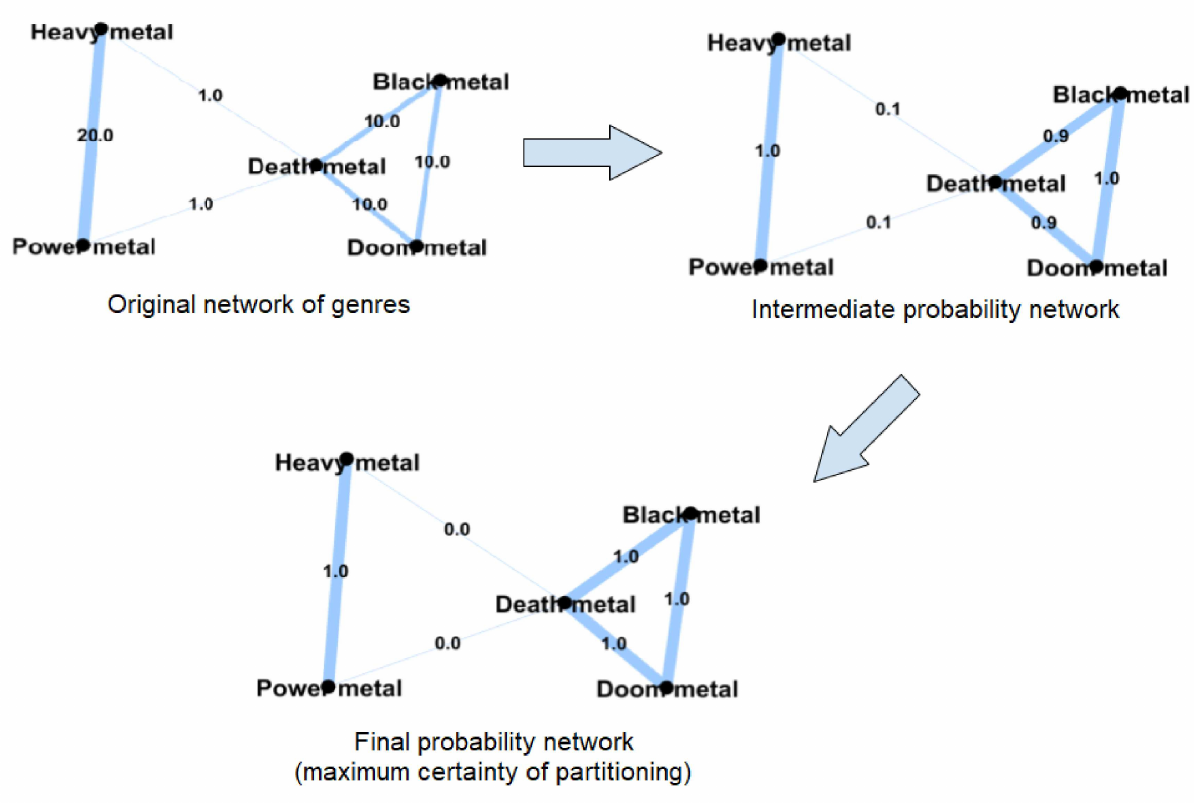}
\caption{\label{fig:average}Process of determining ``average'' partitioning of a genre network using community detection. We start off with the original network of genres in Fig. \ref{fig:bipartite}b, and perform the community detection algorithm multiple times using the original weights of the network to obtain the probabilities that two genres fall in the same cluster, which serve as weight in the new network. We then repeat this process until we obtain maximum certainty, i.e. when the probabilities converge to ones and zeros.}
\end{figure}

The Louvain method for community detection is not definite as exact modularity optimisation would be NP-hard \cite{blondel_fast_2008}. As a result, every run of the algorithm would produce a partition of the network into clusters with slight variation. To obtain stable clusters, we can perform an ``averaged'' partitioning of the genres in the network (see Fig. \ref{fig:average}). Firstly, the network is partitioned using the community detection algorithm for, say 100 times, which generates 100 possible ways to partition the network. For each pair of nodes (i.e. genres), we compute the number of times that these two genres end up in the same cluster. This will be an indicator of the probability that two genres will fall in the same cluster.

To create an ``averaged'' network of genres, a new network of genres is created with the same nodes but the edge weights are now computed such that the weight of an edge between genre A and genre B is the number of times that genre A and genre B fall in the same cluster among the 100 ways of network partitioning. A greater edge weight will indicate that two genres are similar and are likely to be clustered together. This new probability network of genres undergoes another round of 100 times of partitioning, before a new probability network is generated. This process repeats itself until the probabilities (edge weights) converge to 1 or 0. 

In addition, we can also compare the number of edges in the network with the sum of maximum number of edges in each connected component, which is represented by $\epsilon_{\text{max}}$. Equation (\ref{eqn:e_max}) below shows the formula used for $\epsilon_{\text{max}}$
\begin{equation}
\label{eqn:e_max}
\epsilon_{\text{max}} = \sum_{i=1}^{M} \frac{N_i(N_i-1)}{2} \text{,}
\end{equation}
in which $N_i$ represents the total number of nodes in connected component $i$ and $M$ represents the total number of connected components.

When the number of edges in the network has approached $\epsilon_{\text{max}}$, this signals that we can now partition the network with maximum certainty, eliminating all inter-cluster edges such that each cluster will be a maximally connected graph. In other words, each resulting connected component is a separate genre community. Hence we can obtain the average clustering of the genres in the network, to obtain a more reliable partitioning from the community detection algorithm. The convergence of the number of edges and $\epsilon_{\text{max}}$ is illustrated through Table \ref{tab:community}. The number of components refers to the total number of connected components (i.e. connected subgraphs) in the network. The number of edges refer to the total number of edges present in the network. $\epsilon_{\text{max}}$ refers to the sum of maximum number of edges in each connected component, as elaborated on earlier, and given by Eq. (\ref{eqn:e_max}).

\begin{table}[t!]
\caption{Network statistics from each iteration of community detection. See Eq. (\ref{eqn:e_max}) for the calculation of $\epsilon_{\text{max}}$.} 
\label{tab:community}
\centering 
\begin{tabular}{c c c c} 
\hline\hline 
Round  & No. of components & No. of edges & $\epsilon_{\text{max}}$ \\ [0.5ex] 
\hline 
0 (initial) & 1 & 5649 & 5886 \\ 
1 & 1 & 4224 & 5886 \\
2 & 1 & 4224 & 5886 \\
3 & 3 & 2411 & 2411 \\[1ex] 
\hline 
\end{tabular}
\end{table}

\subsubsection{Further breaking down of communities}
\begin{figure}
\centering
\includegraphics[width=\textwidth]{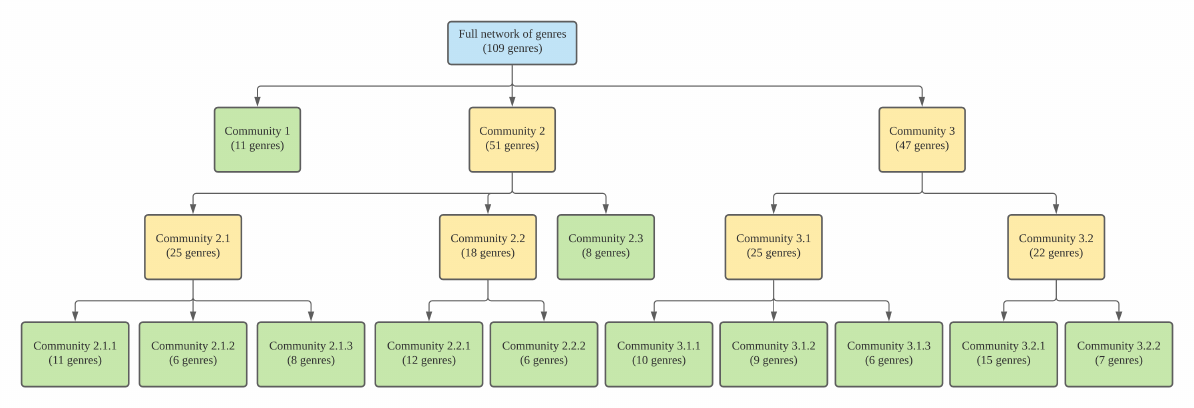}
\caption{\label{fig:clus_stats}Genre communities after each round of community detection (colours online). Genre communities in yellow indicate intermediate communities, and genre communities in green show final communities with no further partitioning required.}
\end{figure}

From a single application of the community detection algorithm, we obtained one small communities (of size 11) and two large communities (size 51 and 47). Thereafter, the two larger communities were further broken down using the same process as described above, in order to obtain smaller communities. In total, 3 layers of community detection were implemented on larger clusters with more genres, as larger clusters contained more genres, making them more difficult to analyse. When deciding which clusters to further break down, we analyse whether the average weight of that cluster has improved from the previous round of clustering. We also took into account the relative size of clusters as compared to other clusters in the network. The details of the different layers of community detection can be found in Fig. \ref{fig:clus_stats}.

\subsubsection{Extracting features from review text}
In order to analyse the unique characteristics of each metal music genre cluster obtained, features need to be extracted from review text written by the users of the genres in each cluster. For this task, chunking methods is often used in product feature extraction of reviews \cite{htay_extracting_2013}, which could, however, be limited as it depends on the position of words in a sentence. Using this concept, after using a Part-of-Speech tagger, nouns and their respective adjectives are extracted using dependency parsing, by establishing relationships between nouns and verbs, offering much greater flexibility in sentence structure and pattern recognition. A ``feature'' is thereafter defined to be an adjective-noun combination. Predetermined dependency structures were used to identify features, to detect both cases when the feature is either the subject or the object of a sentence. The details of the dependency structures used can be found in Table \ref{tab:featuretable}. In the first case, ``nsubj'' refers to the nominal subject, which refers the word which is the subject in the sentence. ``acomp'' is the adjectival complement of a verb. Often, the adjective and noun in a feature is related to each other by a verb. For example, in the sentence ``This album sounds awesome.'', the adjective, ``awesome'', and the noun, ``album'', is separated by the verb ``sounds''. Hence we can use both the ``nsubj'' and ``acomp'' relations to extract the adjective-noun combination from the sentence as the feature. In the second case, it is more direct, as ``amod'' refers to the adjectival modifier of a noun (i.e. the relationship between an adjective and the noun it is describing in a sentence). For example, in the sentence, ``This album has a lyrical theme'', ``lyrical theme'' would be identified as the adjective-noun combination in the sentence. 

\begin{table}
\caption{Dependency tree structures used for adjective-noun feature extraction. ``amod'' refers to the adjectival modifier of a noun, ``nsubj'' the nominal subject, and ``acomp'' the adjectival complement of a verb. See text for more details.} 
\label{tab:featuretable}
\centering 
\begin{tabular}{c c c} 
\hline\hline 
Parent & Dependency & Child/children  \\ [0.5ex] 
\hline 
Noun & amod & Adjective \\ 
- & nsubj and acomp & Adjective and noun \\ [1ex] 
\hline 
\end{tabular}
\end{table}

Based on these patterns, we paired up nouns and their related adjectives into features for each set of review. For each feature, we performed lemmatisation to find the root word for each of the adjectives and nouns, to remove duplicates of features due to different word forms (such as singular or plural nouns). Based on the clusters identified earlier, we obtained a list of features for each genre in every cluster. Based on the list of features for each cluster, we calculate the frequency at which each feature appears for each cluster based on the genres in the cluster. This will be the term frequency, which will be elaborated on in the next section.

\subsubsection{Determining most important features}
Term Frequency - Inverse Document Frequency (TF-IDF) \cite{ramos_using_2003} is performed on each feature for each cluster. A ``document'' is defined as a cluster of genres. The Term Frequency (TF) is defined as the frequency that a feature is mentioned within a cluster. The Inverse Document Frequency (IDF) is inversely related the number of clusters in which the feature appears. The TF-IDF is the product of TF and IDF, as shown the formula below

\begin{equation}
tfidf = tf \times idf \text{.}
\end{equation}

The TF accounts for the frequency at which the feature appears, which corresponds to its relative importance in the dataset. In other words, the higher the TF, the more important the feature is in the cluster. The IDF ensures that the features extracted have a certain degree of uniqueness to their group (or document), in order to remove common features which are not representative of the document, which in our case, refers to a cluster of genres. Features which are more unique to one cluster, and not found in other clusters, have a higher IDF. In essence, the IDF imposes a penalty on the TF, based on how common (or uncommon) the feature is across different clusters.

The IDF was modified to be applied on the adjective only in the two-word feature, in order to filter out general adjectives such as ``great'', ``awesome'' and ``amazing'', which add little value to describe the clusters, instead of applying the filter to the whole feature which is often more specific. This step is performed as we observed that by doing so, we can eliminate more general descriptions of each cluster, and effectively extract the more specific characteristics of each cluster instead. The top features of each cluster can be obtained based on the highest TF-IDF to describe each cluster. For practical reasons, we chose the top 50 features based on TF-IDF for analysis. 

The advantage of the method described above is that it can eliminate generic descriptions in the reviews, which were about how much the reviewer enjoyed (or hated) the music instead of the quality of music. These emotive phrases often repeat across different genres and clusters, and can be effectively filtered out by the TF-IDF. Hence we will be left with more features which are specifically describing the characteristics of particular clusters. For the cluster analysis in the following sections, only the top 50 features per cluster based on TF-IDF were considered. Based on our evaluation, we found that the accuracy of feature extraction, is slightly above 70\%, for the top 50 features of each cluster.  This accuracy is calculated by the following equation
\begin{equation}
\label{acc}
\text{accuracy} = \frac{n_{\text{correct}}}{n_{\text{total}}} \times 100\%\text{,}
\end{equation}
in which $n_{\text{correct}}$ represents the number of features correctly identified per cluster, and $n_{\text{total}}$ represents the total number of features extracted from that cluster (which is 50).

\section{Results and Discussion}

\begin{figure}[h!]
\centering
\includegraphics[width=\textwidth]{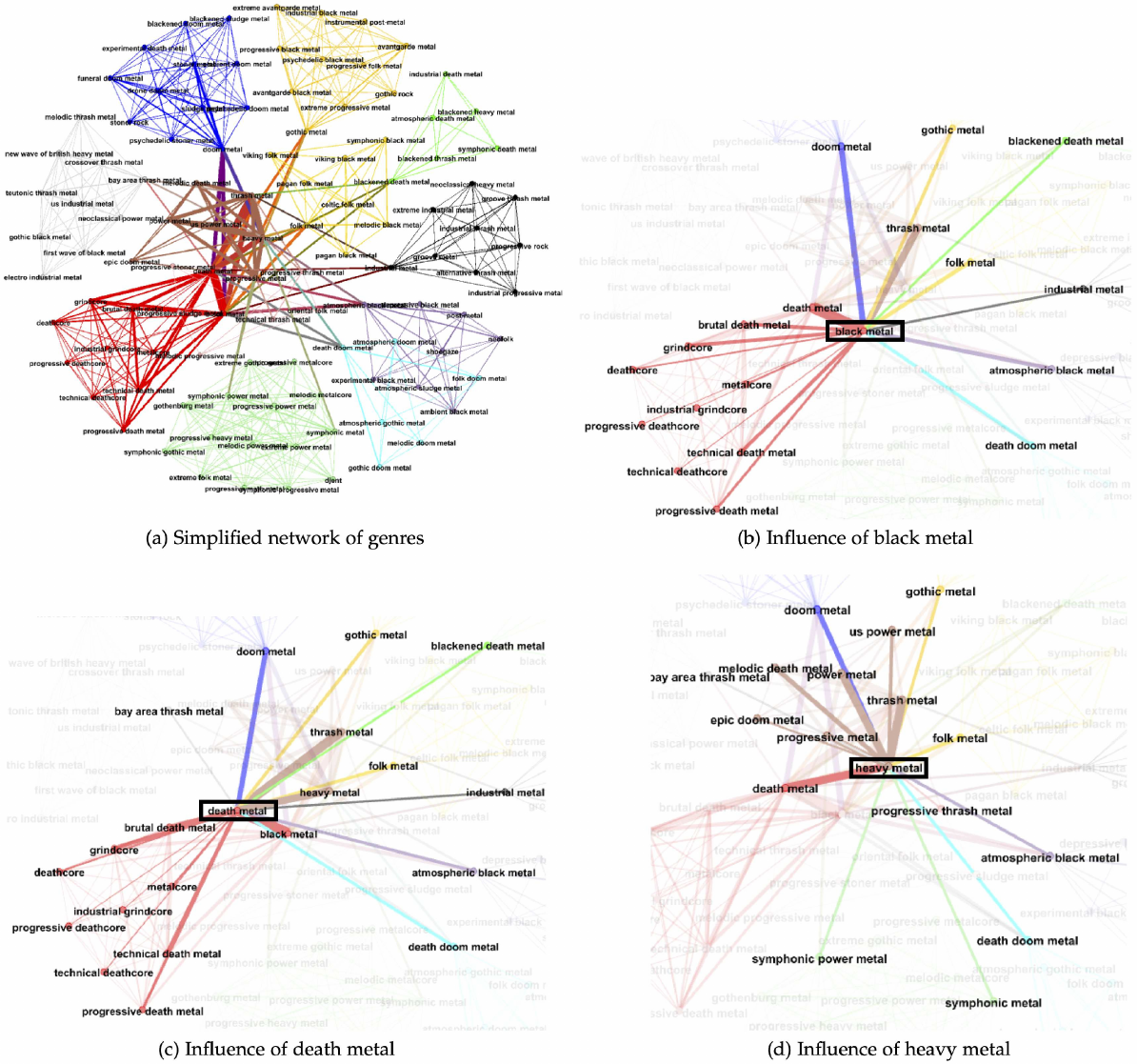}
\caption{\label{fig:three graphs}Network of genres (colours online). Colours of edges represent different clusters identified. In (b), (c) and (d), only the central genres and its neighbours are highlighted.}
\end{figure}

In this study, a total of 12 clusters were obtained from the clustering process, varying in size from 6 to 15 nodes (see Fig. \ref{fig:clus_stats}). Fig. \ref{fig:three graphs}a shows the full network of genres, where different clusters are represented by different colours, and the width of an edge is determined by its weight. For the figures of individual clusters in Fig. \ref{fig:genrecomms}, the width and colour intensity of the edges in the figures are representative of their weight. While all intra-cluster edges are shown in Fig. \ref{fig:three graphs}a, for inter-cluster edges, only the 3 out-edges from the cluster which have the highest edge weights are plotted. This is to better observe the interaction between different clusters, which was a similar visualisation method used by \cite{silver_genre_2016}. This visualisation method allows us to observe the main cross-community edges coming out of each cluster, in order to better visualise the interactions between different clusters, and identify the main core genres in the network. The graph was drawn using force-directed graph drawing (Fruchterman Reingold algorithm) developed by \cite{fruchterman_graph_1991}. All graph visualisations were generated using Gephi, a graph visualisation software \cite{bastian_gephi_2009}. 

\subsection{Genre hierarchy at the network level}

There are 3 central genres in the constructed network. The first two central genres are the black and death metal. As shown in Fig. \ref{fig:three graphs}b and \ref{fig:three graphs}c, these genres are not only connected within their own cluster (represented in red), but also are connected to a wide variety of other genres from different clusters (represented by different edge colours). The presence of inter cluster edges show that these were the original (and overarching) genres, which further split into different subgenres of black and death metal over time. For example, in the case of death metal, as shown in Fig. \ref{fig:three graphs}b, it strongly influences its subgenres and derivatives such as blackened death metal and death doom metal, from outside the cluster. It also has some influence on non-death metals such as doom metal, gothic metal and industrial metal. This shows that the genre has significant influence even outside its own cluster, which makes the genre more important in the network. While not all of these subgenres fall in the same cluster as black and death metal, it is clear that black and death metal have strong connections other genres and genre clusters in the network. 

Similarly, we also observe such pattern for traditional heavy metal. Traditional heavy metal music (henceforth referred to simply as ``heavy metal''\footnote{While ``heavy metal'' (or simply ``metal'') is generally known as a genre of rock music, it is also a name of a subgenre within metal, which features pioneering bands such as Iron Maiden or Judas Priest. In this work, we reserve the term ``heavy metal'' for the subgenre, as used in the MA and MS databases.}) is one of the earliest and most original forms of metal music, hence we can see its influence on a wide variety of other genres as well (Fig. \ref{fig:three graphs}d). The most notable observation would be that it has strong influences to different forms of power metal, which originate from heavy metal and contain some elements of heavy metal. In fact, the cluster that heavy metal belongs to, represented in brown in Fig. \ref{fig:three graphs}a, shows that other important genres, such as progressive metal, power metal and thrash metal, are all influenced strongly by heavy metal. This may explain the central position of this cluster of main genres in the entire network of genres in Fig. \ref{fig:three graphs}a. Hence we can conclude that the genre network after community detection shows certain hierarchical patterns, as main genres (black, death and heavy metal) branch out into smaller subgenres over time, as genres and user preferences become more diverse. 

\begin{figure}[h!]
\centering
\includegraphics[width=0.9\textwidth]{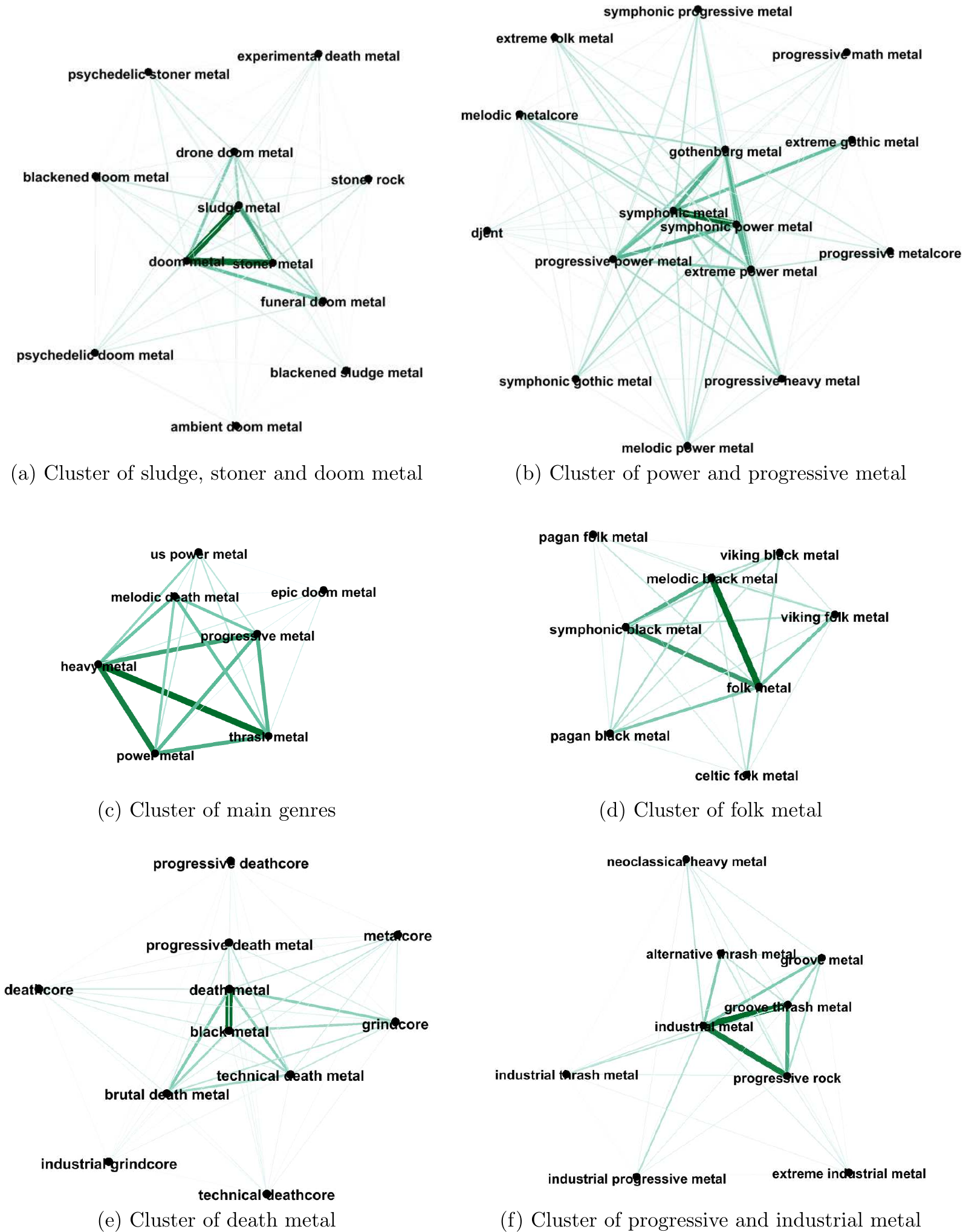}
\caption{\label{fig:genrecomms}Selected genre communities.}
\end{figure}

This network representation of genres also allows us to easily identify the principal genre(s), or the most important genres, which are representative of each cluster. The 3 out-edges from each cluster with greatest edge weight often only originate from one or two specific genres, which are often the principal genres of that particular cluster. An example is the cluster in blue (also shown in Fig. \ref{fig:genrecomms}a), which has the principal genre of doom metal, and all the 3 out-edges originate from the doom metal node. The principal genre detected using this method is often the main genre of the cluster which tend to encompass more subgenres, hence the greater number and strength of connections to other clusters due to overlapping (common) characteristics and user interest between different clusters.

\subsection{Nature of music}

Genres are observed to be grouped according to the nature of music, based on user interests. As shown in Fig. \ref{fig:genrecomms}b, different types of progressive\footnote{\label{noauthor_progressive_nodate-1}\url{http://www.progarchives.com/subgenre.asp?style=19}} and power metal are closely associated together and share similar nature of music, both originating from progressive rock and heavy metal\footnote{\label{bowar_quick_2019}\url{https://www.liveabout.com/what-is-heavy-metal-1756179}}. Both progressive and power metal contain some elements of heavy metal. Similarly, progressive metal was popularised by bands which were playing power and heavy metal (e.g. Queensrÿche and Fates Warning)\footnote{\url{https://metal.fandom.com/wiki/Progressive_metal}}. Hence, it is natural that these genres would be enjoyed by a similar community of users, resulting in their being clustered together. Some of the main features extracted for this cluster include ``harsh vocal'', ``clean vocal'' and ``male vocal'', suggesting that vocals play an important role in this cluster, which is accurate as power metal places high emphasis on the vocalist. In fact, we also detected the feature ``soaring vocal'', which is a distinct characteristic of power metal as it involves soaring guitars and vocals\footref{bowar_quick_2019}. The feature selection also picked up ``catchy chorus'' which is a common characteristic of power and progressive metal, often featuring rhythmic and easy-to-remember choruses. ``Lyrical theme'' and ``lyrical content'' were also some of the top features extracted from review text, as these features were important in shaping the progressive and power metal genres\footref{noauthor_progressive_nodate-1}. Modern progressive music also incorporated more rhythmic elements such as syncopation\footref{noauthor_progressive_nodate-1}, and this is also reflected in the reviews in features such as ``rhythm section'' and ``rhythm guitar''. 

Symphonic metal was also featured prominently in this cluster. Symphonic metal bears many similarities with progressive metal as progressive metal often draws influence from symphonic music\footref{noauthor_progressive_nodate-1}. Symphonic metal is also very similar to power metal, with the addition of more classical elements in the music\footnote{\label{noauthor_symphonic_nodate}\url{https://metal.fandom.com/wiki/Symphonic_metal}}. Hence these 3 types of metal music genres are clustered together in the same community. This also explains why ``female vocal'', ``orchestral arrangement'' and ``classical music'' were also part of the top features extracted, as symphonic metal bands often showcase female vocalists with classical backgrounds, who often have lead roles in the music. ``Operatic vocal'' was also detected, explaining why the nickname for symphonic metal is ``opera metal'', as it often features operatic themes. The feature extraction also captured the time period of ``late 90s'' for this particular cluster. This is because it was during that time period (1997) that bands such as Nightwish and Within Temptation, who are some of the pioneer bands of symphonic metal, first released their symphonic metal albums and started gaining popularity. 

Based on the cluster of main genres (Fig. \ref{fig:genrecomms}c), which is a strongly connected cluster comprising of some of the main genres, we can observe a similar trend that heavy metal, power metal, progressive metal and thrash metal are part of the same common interest community as they have similar origins. This cluster was originally from the same genre community as the cluster in Fig. \ref{fig:genrecomms}b, and this network core was extracted as a separate genre community in later stages of the genre community detection. The original genre community was community 3.2 in Fig. \ref{fig:clus_stats}. This further supports our core-periphery theory for genre communities, which will be elaborated on in Section \ref{structure}.

\subsection{Geographical patterns}

The cluster shown in Fig. \ref{fig:genrecomms}d shows that a certain group of users generally prefer Viking, Celtic and Pagan metal, which are closely related to folk metal. Folk metal is a form of music which is often inspired by Paganism, and folklore\footnote{\url{https://www.villagevoice.com/2013/12/18/the-10-best-folk-metal-bands/}}. The geographical significance is reflected in our review text feature extraction, where the top features of this cluster included  ``Norwegian band'', ``Finnish band'', ``Swedish band'', ``Norwegian scene''. This is because Viking metal and pagan metal both originated from the Nordic countries, and have themes stemming from Norse Mythology and Viking history\footnote{\url{https://metal.fandom.com/wiki/Viking_metal}}. The first Viking metal album, Hammerheart (1990) was released by the band Bathory from Sweden, as it was solely based on Norse Mythology\footnote{\url{https://metalinsider.net/today-in-metal/today-in-metal-hammerheart-by-bathory-turns-24-years-old}}. In the 1990s, In the Woods... from Norway were one of the first bands to label their music as pagan metal, and later on developed on the genre by defining genre characteristics like using traditional instruments\footnote{\url{https://www.metal-archives.com/bands/In_the_Woods.../392}}. Examples of folk instruments used are fiddles, flutes and bag pipes\footnote{\url{https://www.villagevoice.com/2013/12/18/the-10-best-folk-metal-bands/}}. 

The heavy Nordic influence in this genre community can also be observed in terms of the number of reviews of albums which are tagged with genres in the folk metal cluster, as shown in Table \ref{tab:geopatterns}. The three Scandinavian countries, Norway, Finland and Sweden, are observed to be the most popular countries of origin for this particular cluster, which is consistent which our earlier findings from the analysis of review text. In particular, out of the 2,928 reviews for the albums of genres in the folk metal cluster (see Fig. \ref{fig:genrecomms}d), almost a fifth are for the ones released by bands from Norway. Together with fellow bands from Finland and Sweden, they account for two fifths of such reviews. Extending to neighbouring countries like the UK and Germany, the corresponding bands receive close to 60 per cent of the reviews.
{\begin{table}[t]
\caption{Top countries of origin for bands whose releases are tagged with genres in the folk metal cluster (see Fig. \ref{fig:genrecomms}d).} 
\label{tab:geopatterns}
\centering 
\begin{tabular}{c c} 
\hline\hline 
Country of origin & Number of reviews\\ [0.5ex] 
\hline 
Norway & 507 (17.3\%) \\ 
Finland & 343 (11.7\%)\\
Sweden & 322 (11.0\%)\\
United Kingdom & 275 (9.4\%)\\
Germany & 271 (9.3\%)\\ [1ex] 
All countries & 2,928 \\
\hline 
\end{tabular}
\end{table}}

The lyrics in these genres originating from folk metal often have reference to paganism, nature, fantasy, mythology and history\footnote{\url{https://metal.fandom.com/wiki/Folk_metal}}. Some features unique to this cluster include ``native language'', ``native tongue'', ``traditional instrument'', ``Celtic music'', ``Celtic influence'' and ``folkish melody'', which are present in folk metal as elaborated on earlier. It is also consistent with the fact that many bands choose to perform in their native languages. This shows that metal music fans are looking out for these characteristics in folk metal music, as these are the features which make folk metal stand out from other genres, and these characteristics have strong correlations to the geographical origins of these folk metal bands due to the strong reference to Nordic, Viking and Celtic history and culture. Hence, it is clear that user interest in metal music can be categorised by geographical proximity of bands. 

\subsection{Structure of network}
\label{structure}
Similar to Fig. \ref{fig:genrecomms}b where genres in the same cluster often share similar properties, genres related to death metal are also in the same cluster as shown in Fig. \ref{fig:genrecomms}e. There are two parts of this network, the core and the periphery. At the core, there are two of the main overarching genres, black and death metal, with extremely high edge weight. In the second layer, we see the subgenres emerging, which are brutal, progressive and technical death metal. At the periphery (third layer), genres which contain the suffix ``-core'' were more common. Metalcore is a combination of extreme metal and hardcore punk, and deathcore is a combination of death metal and metalcore\footnote{\url{https://theydiffer.com/difference-between-metalcore-death-metal-and-deathcore/}}. Hence, this trend can be due to the perception that these genres are not very metal in nature, and as a result, less popular within the mainstream metal music community. In addition, black and death metal are the principal genres in this genre cluster, hence the other genres are subgenres which were derived from black and death metal, which explains why the different layers seem to branch out from the center, with edge weight decreasing as the nodes are further from the core containing black and death metal. As elaborated on earlier, these genres often evolve over time and develop into further subgenres and become more diverse, which is evident in the layering effect of this genre community. This pattern of multi-core genre networks is evident in other clusters obtained (e.g. cluster of doom, sludge and stoner metal in Fig. \ref{fig:genrecomms}a), and also other types of genre networks \cite{silver_genre_2016}. Our results also show that this cluster exhibits layering properties beyond just 2 layers observed in prior research.

In addition, with reference to the black and death metal cluster in Fig. \ref{fig:genrecomms}e, the features extracted also reflect the diversity of time periods as the black and death metal genres evolve. We observe that some of the top features in the review text were ``late 80s'', ``early 90s'' and ``21st century'', which shows the time span encompassed by this genre community. Death, one of the first death metal bands, released their first album in 1987. Cannibal Corpse, another popular American death metal band was formed in Buffalo, New York in 1988\footnote{\url{https://loudwire.com/heavy-metal-101-history-of-death-metal/}}. Hence it is evident that death metal started to gain traction especially in America during the late 80s and early 90s. While death metal saw its popularity rise and fall, in the mid-90s, death metal suddenly became popular again due to bands such as Entombed and At The Gates, who brought a distinct European flavour to the genre\footnote{\url{https://www.loudersound.com/features/the-50-best-death-metal-albums-ever}}. Toward the end of the decade, another wave of death metal arrived, led by bands from Sweden like Dark Tranquillity or In Flames who combined old-school guitar harmonies with the harsh style of death metal\footnote{\url{https://www.metalcrypt.com/genres.php}}. The subgenre is later known as melodic death metal for the injection of harmonic guitar work (or Gothenburg metal for the origin of the pioneering bands) and continues to be a strong force in metal at the turn of the century and beyond.

From this analysis, we can conclude that the genre network not only exhibits hierarchical patterns on a network level where certain key genres branch out into subgenres from many different genre communities, but also at a cluster level, where we can easily identify the main genres of genre communities and observe how these genres evolve as they become more diverse. The features extracted from the review text also shows the period in which genre communities evolve, adding further contextual knowledge to the evolution of genre communities.

\section{Conclusion}

Overall, the results in this study indicate that related genres can be identified and clustered together through common-interest communities. The user-oriented nature of our network representation and community detection was greatly useful in clustering common user interests, encapsulating the idea that genre structures are subjective and should be determined by user preferences. Hence, the network of genres can be used to analyse the music genre structure due to the presence of common-interest communities. The network of genres exhibits unique structures which allow us to understand the relationships between different genres, and different clusters of genres. Based on these genre communities, we can conclude that genre hierarchy is present between different genres. At the network level, we observe certain core genres in the network with a high number of cross-community edges, which shows that these genres were the original metal music genres, which further branch out into subgenres over time. At a cluster level, we also observe similar layering patterns which represent genre hierarchy. This method is useful in detecting the main genres in a dataset, based on the influence on its subgenres via the network, as well as how well it represents its genre community as a principal genre of that community. Hence a network offers many more insights on the characteristics of these genres as compared to other statistical methods. 

Genres were also clustered together based on a variety of factors which influence the musical tastes of metal music fans, such as geography and nature of music. These factors were captured in our review text analysis effectively, by using Dependency Parsing and TF-IDF to extract review features, which offered distinctive ways to describe different clusters, complementing the structural characteristics of the cluster. This shows the huge potential of review text analysis and extraction in describing genre communities, beyond basic quantitative measures like average weight and clustering coefficient, which have limited effectiveness in dense genre networks.

The classification method developed in this work can be useful in determining the tags or labels of specific genres and albums, where it can work as folksonomy to help users determine the grouping of the items instead of the moderators of a platform. This form of classification is more organic and objective, with the potential to tackle larger datasets where tagging can be a problem. In addition, our methodology can also be used to help understand why communities of common interest are formed by people, by tapping into the structural and textual hints extracted from the community. This can be helpful in recommendation systems for users who are seeking out similar genres on the platform, and help them to discover their specific interests over time as they move along the genre hierarchy. In addition, musicians can also use these genre structures to discover their own niche, as well as tag their music with the appropriate labels to attract suitable audiences for their music in the common-interest community \cite{silver_genre_2016}. Such classification methods can also be applicable beyond music platforms. For example, a similar network method can be used to classify different types of products with users' reviews on platforms such as Amazon, and analyse common-interest communities to account for the popularity of certain groups of products when it comes to online shopping.

For future work, given a sufficiently large dataset, with some slight modifications to our method, the effects of time on the relationships between different genres (or categories) can also be studied, to further illustrate the temporal evolution of the network of genres over time. This can allow us to observe the rise and fall in popularity of certain genres types over time, as well as provide insight into how these genre structures are formed. In addition, to further investigate how the hierarchical structure of the network is created, we can make use of the active periods of different bands to consider how older bands might influence newer ones to create derivative genres. This for example could be done by studying the diversification rate (i.e. birth and death) \cite{koch_evolutionary_2020} of cultural objects like genres and study the changes in quantity, types and quality of cultural ideas over time.

%
%
%

\section*{Acknowledgments}

ZJ would like to thank Hwa Chong Institution's research unit for their support.

\bibliography{references}


\end{document}